\documentclass[3p,times,procedia]{elsarticle}
\usepackage{ecrc}

\volume{00}

\firstpage{1}

\journalname{Nuclear Physics A}

\runauth{D. Kiko\l{}a for the AFTER@LHC study group }

\jid{nupha}

\jnltitlelogo{Nuclear Physics A}

\usepackage{amssymb}

\usepackage[figuresright]{rotating}
\usepackage{subfigure}

\newcommand{\sqrtsNN }{\mbox{$\sqrt{s_{\rm NN}}$}}
\def\sqrts {\mbox{$\sqrt{s}$}}
\def\AFTER {\mbox{AFTER@LHC}}

\def\pp   {pp}

\def\pA   {pA}

\def\AA   {AA}

\def\PbPb {PbPb}
\def\PbA {PbA}
\def\PbXe {PbXe}

\def\pA {pA}

\def\sqrts {\mbox{$\sqrt{s}$}}
\newcommand{\sqrtS}[1]{\mbox{$\sqrt{s_{#1}}$}}

\newcommand{\eg}{{\it e.g.}}

\newcommand{\calL}{\!\mathcal{L}}
\def\jpsi    {\mbox{$J/\psi$}}

\def\jpsi    {\mbox{$J/\psi$}}

\usepackage[colorlinks=true,linkcolor=somecolor,citecolor=somecolor]{hyperref}

\begin{document}
\begin{frontmatter}

\dochead{XXVIIth International Conference on Ultrarelativistic Nucleus-Nucleus Collisions\\ (Quark Matter 2018)}

\title{A fixed-target programme at the LHC for heavy-ion, hadron, spin and astroparticle physics: AFTER@LHC}

\author[WUT]{D.~Kiko\l{}a\corref{dk}}
\address[WUT]{Faculty of Physics, Warsaw University of Technology, ul. Koszykowa 75, 00-662 Warsaw, Poland}
\ead{daniel.kikola@pw.edu.pl}

\cortext[dk]{Corresponding author}

\author[SLAC]{S.J.~Brodsky} 
\address[SLAC]{SLAC National Accelerator Laboratory, Stanford University, Menlo Park, CA 94025, USA}

\author[Roma]{G.~Cavoto} 
\address[Roma]{``Sapienza" Universit\`a di Roma, Dipartimento di Fisica \&
INFN, Sez. di Roma, P.le A. Moro 2, 00185 Roma, Italy}

\author[LANL]{C.~Da~Silva}
\address[LANL]{P-25, Los Alamos National Laboratory, Los Alamos, NM 87545, USA}

\author[Turin]{F.~Donato}
\address[Turin]{Turin University, Department of Physics, and INFN, Sezione of Turin, Turin, Italy}

\author[Pavia]{M.G.~Echevarria}
\address[Pavia]{Istituto Nazionale di Fisica Nucleare, Sezione di Pavia, via Bassi 6, 27100 Pavia, Italy}

\author[USC,LLR]{E.G.~Ferreiro}
\address[USC]{Dept. de F{\'\i}sica de Part{\'\i}culas \& IGFAE, USC, 15782 Santiago de Compostela, Spain}
\address[LLR]{Laboratoire Leprince-Ringuet, Ecole polytechnique, CNRS/IN2P3, Universit\'e Paris-Saclay, Palaiseau, France}

\author[IPNO]{C.~Hadjidakis}
\address[IPNO]{IPNO, CNRS-IN2P3, Univ. Paris-Sud, Universit\'e Paris-Saclay, 91406 Orsay Cedex, France}

\author[IPNO]{I.~H\v{r}ivn\'{a}\v{c}ov\'{a}}

\author[LANL]{A.~Klein}

\author[INR]{A.~Kurepin}
\address[INR]{Institute for Nuclear Research, Moscow, Russia}                    

\author[Krakow]{A.~Kusina}
\address[Krakow]{Institute of Nuclear Physics Polish Academy of Sciences, PL-31342 Krakow, Poland}

\author[IPNO]{J.P.~Lansberg}

\author[CPHT]{C.~Lorc\'e}
\address[CPHT]{CPHT, \'Ecole Polytechnique, CNRS,  91128 Palaiseau, France}

\author[SMU]{F.~Lyonnet}
\address[SMU]{Southern Methodist University, Dallas, TX 75275, USA}

\author[BNL]{Y.~Makdisi}
\address[BNL]{Brookhaven National Laboratory, Collider Accelerator Department}

\author[IPNO]{L.~Massacrier} 

\author[LPC]{S.~Porteboeuf}
\address[LPC]{Universit\'e Clermont Auvergne, CNRS/IN2P3, LPC, 63000 Clermont-Ferrand, France.}

\author[LIP]{C.~Quintans}
\address[LIP]{LIP, Av. Prof. Gama Pinto, 2, 1649-003 Lisboa, Portugal}

\author[DPhN]{A.~Rakotozafindrabe} 
\address[DPhN]{IRFU/DPhN, CEA Saclay, 91191 Gif-sur-Yvette Cedex, France}

\author[LAL]{P.~Robbe}
\address[LAL]{LAL,CNRS-IN2P3, Univ. Paris-Sud, Universit\'e Paris-Saclay, 91898 Orsay Cedex, France}

\author[CERN]{W.~Scandale} 
\address[CERN]{CERN, European Organization for Nuclear Research, 1211 Geneva 23, Switzerland}

\author[LPSC]{I.~Schienbein}
\address[LPSC]{LPSC, Universit\'e Grenoble Alpes, CNRS/IN2P3, 
F-38026 Grenoble, France}

\author[JS]{J.~Seixas}
\address[JS]{Dep. Fisica, Instituto Superior Tecnico \& LIP,  Lisboa, Portugal}

\author[LPTHE]{H.S.~Shao}
\address[LPTHE]{LPTHE, UMR 7589, Sorbonne University\'e et CNRS, 4 place Jussieu, 75252 Paris Cedex 05, France}

\author[JLAB]{A.~Signori}
\address[JLAB]{Theory Center, Thomas Jefferson National Accelerator Facility, Newport News, VA 23606, USA}

\author[INR]{N.~Topilskaya}

\author[Utrecht]{B.~Trzeciak}
\address[Utrecht]{Institute for Subatomic Physics, Utrecht University, Utrecht, The Netherlands}

\author[IPNL]{A.~Uras} 
\address[IPNL]{IPNL, Universit\'e Claude Bernard Lyon-I and CNRS-IN2P3, Villeurbanne, France}

\author[NCBJ]{J.~Wagner}
\address[NCBJ]{National Centre for Nuclear Research (NCBJ), Ho\.{z}a 69, 00-681, Warsaw, Poland}

\author[IPNO]{N.~Yamanaka}

\author[CHEP]{Z.~Yang}
\address[CHEP]{Center for High Energy Physics, Department of Engineering Physics, Tsinghua University, Beijing, China \vspace*{-1.2cm}}

\author[BNL]{A.~Zelenski}

\begin{abstract}
Thanks to its multi-TeV LHC proton and lead beams, the LHC complex allows one to perform the most energetic fixed-target experiments ever and to study with high precision pp, pd and pA collisions at \sqrtsNN = 115 GeV and Pbp and PbA collisions at \sqrtsNN = 72 GeV. We present a selection of feasibility studies for the production of quarkonia, open heavy-flavor mesons as well as light-flavor hadrons in \pA\ and \PbA\ collisions using the LHCb and ALICE detectors in a fixed-target mode.
\end{abstract}


\end{frontmatter}


\section{Introduction}
\label{sec:intro}

In the two last decades, the majority of the efforts and resources in the field of relativistic heavy-ion collisions  has been devoted to collider experiments at the Relativistic Heavy Ion Collider (RHIC) and the Large Hadron Collider (LHC). They have been driven by a quest for higher energy collisions to reproduce, in the laboratory, a state of matter where the degrees of freedom are the quarks and gluons -- the Quark-Gluon Plasma (QGP). Fixed-target experiments using highly energetic beams offer unique features facilitating studies of specific QGP properties, which are inaccessible in a collider setup~\cite{Brodsky:2012vg,Lansberg:2012kf,Kikola:2015lka,Massacrier:2015qba,Trzeciak:2017csa,Kikola:2017hnp,Hadjidakis:2018ifr}. First, the available luminosities are usually much higher than those reachable at the same energy in a collider setup; this gives access to precise studies of rare probes (charm and bottom quarks). Second, a large variety of available target materials allows for a system-size dependent study of the nuclear effects. Third, the wide kinematic coverage of existing collider experiments used in the fixed-target mode gives an additional handle on the longitudinal expansion of the system created in these collisions. Finally, such experiments allow for studies of the structure of the nucleons in a range where  partons carry large \textit{x}-Bjorken fractions of the nucleon momentum, the \textit{high-x} domain. In this paper, we briefly review the physics opportunities offered by a fixed-target program using the proton and ion beams of the LHC, which we dub the \AFTER\ program. We also review the technical possibilities of its realization.

\section{Kinematic features and implementation possibilities}
\label{sec:after}

As just mentioned, \AFTER~\cite{Hadjidakis:2018ifr} is a proposal for a fixed target program at the LHC. Such an endeavor could be realized in a cost-effective way, using existing ALICE and LHCb detectors. These detectors excel in particle identification, offer complementary rapidity coverages and could take data in parallel with an operation in the collider mode. 

The collision of 2.76 TeV lead beam on a fixed target results in a center-of-mass (c.m.s)  energy per nucleon pair of \sqrtsNN = 72 GeV, while 7 TeV proton beam gives \sqrts = 115 GeV. For such energetic beams, there is a large boost between the laboratory and the c.m.s. frames: 4.8 units of rapidity for \pA\ collisions with a proton beam, and 4.3 for a heavy-ion beam. The forward instrumentation of the LHCb and ALICE experiments therefore covers the mid-rapidity in the c.m.s. system, while the ALICE central barrel becomes a far-backward detector. Such large boost is an important asset in this case. It allows for relatively easy measurement of D and B mesons using the displaced vertex technique with the existing apparatus of the LHCb detector. Moreover, one can access the \textit{high-x} range ($x \rightarrow 1$) in the target by selecting tracks in the ALICE central barrel. 

\begin{figure}[!hbt]
\centering
\begin{tabular}{ccc}
\includegraphics[width=0.32\textwidth]{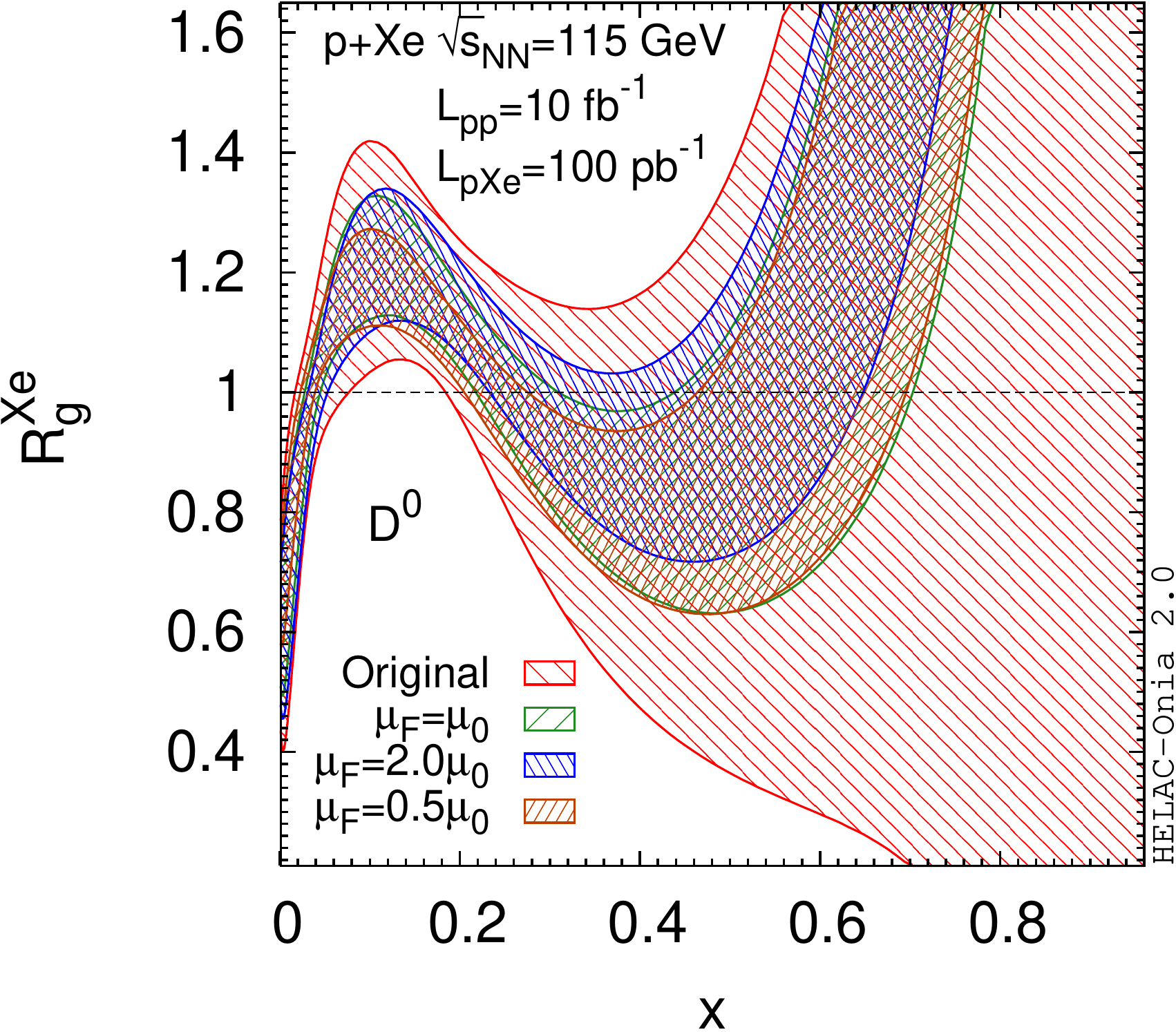} &
\includegraphics[width=0.32\textwidth]{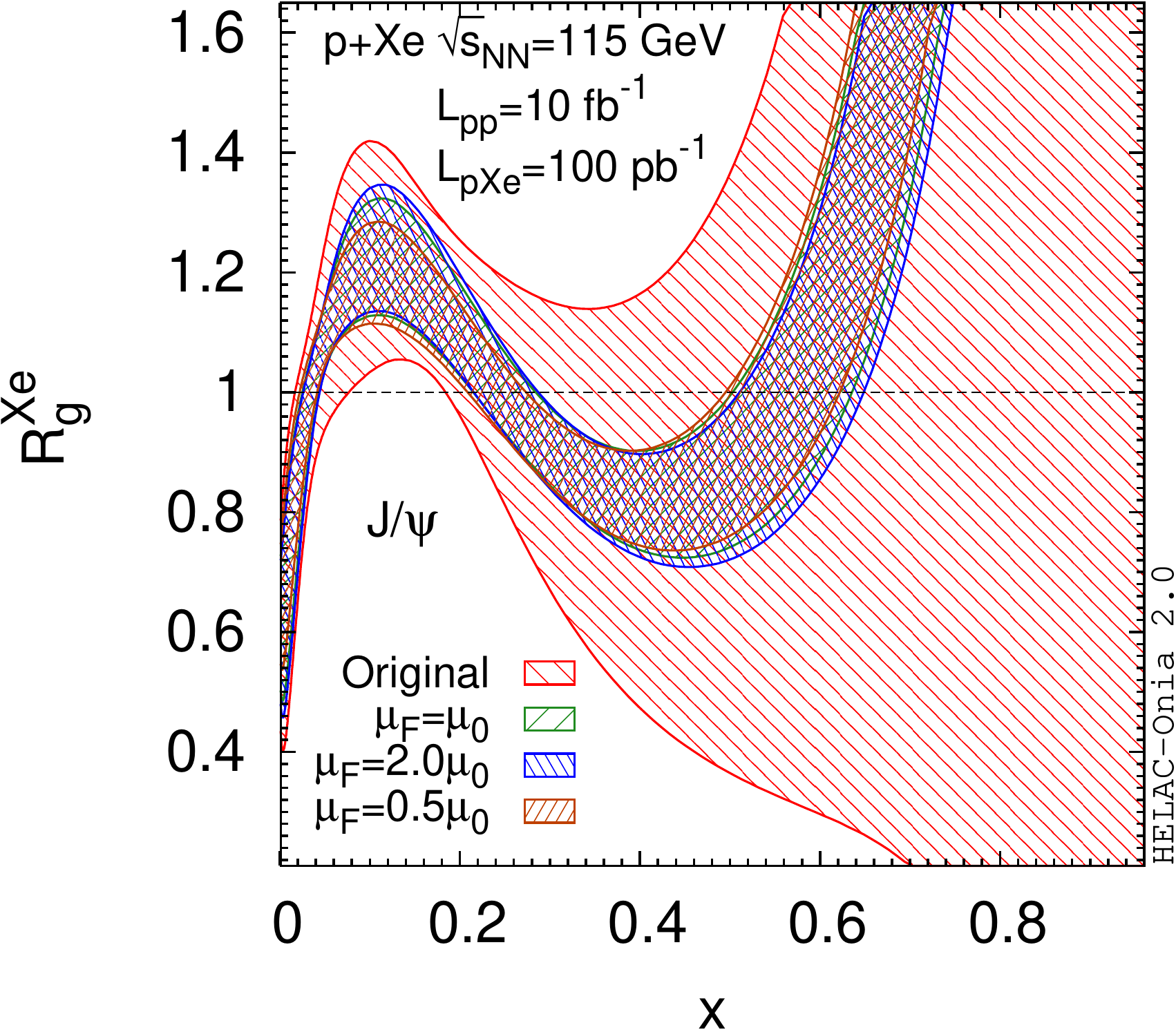}
\includegraphics[width=0.32\textwidth]{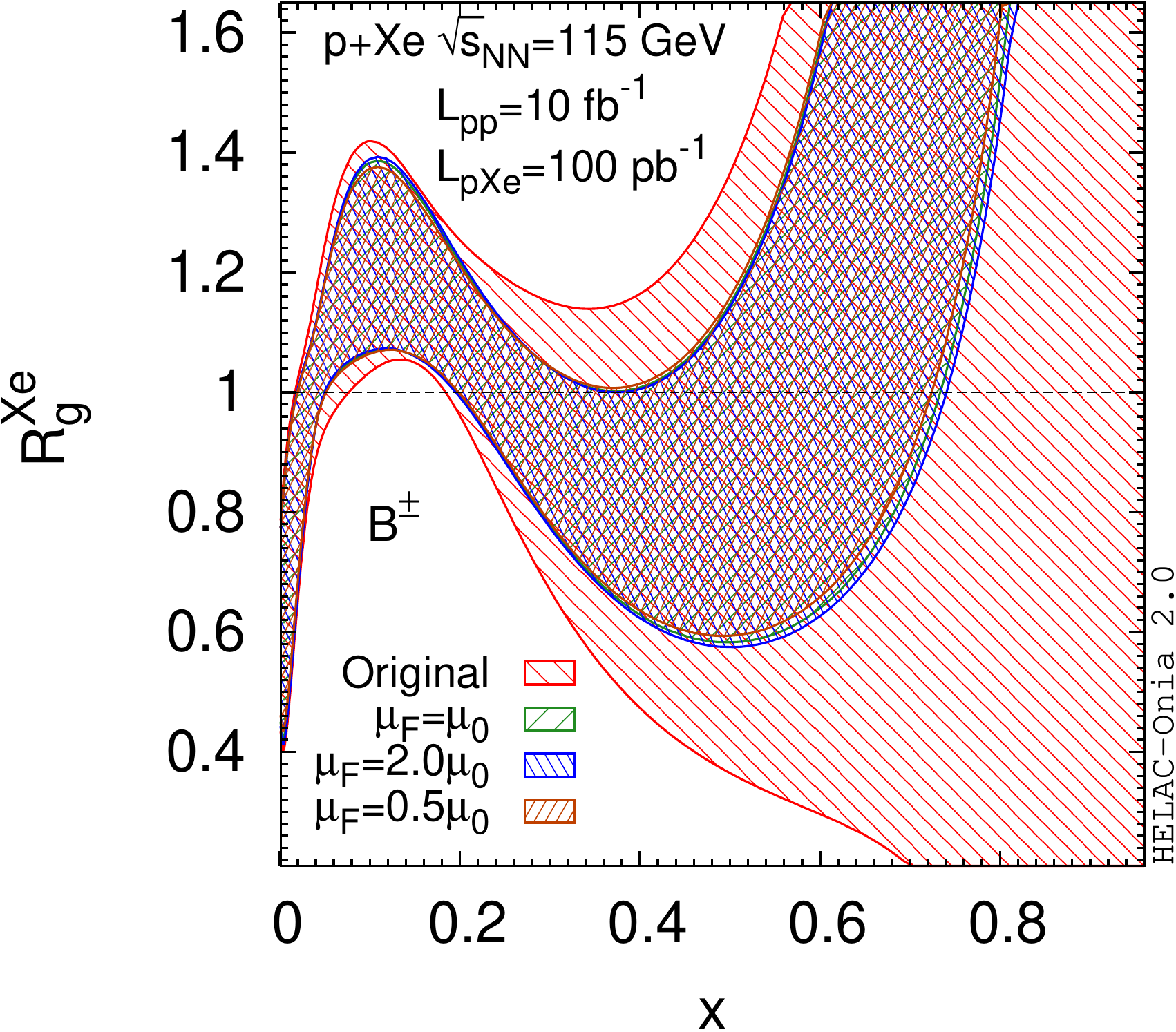}
\end{tabular}
\caption{The improvement of the nuclear gluon density $R_g^{Xe}(x)$ determination expected with the D$^0$, $J/\psi$, B$^+$ production measurements in \AFTER. Figure taken from~\cite{Hadjidakis:2018ifr}.}
\label{fig:gluon_RpA}
\end{figure}

The most feasible options for the realization of the \AFTER\ program is the installation of a fixed target (gaseous or solid) within or next to an existing detector: LHCb or ALICE. A gas target (unpolarized or polarized) could be exposed to the full LHC beam - in such a case we expect an integrated luminosity $\int \calL$ to be up to 10 fb$^{-1}$/year for \pp\ collisions at \sqrtS\ = 115~GeV, $\int \calL \sim 100$~pb$^{-1}$/year for \pA\ reactions at  \sqrtsNN = 115~GeV, and $\int\calL \sim 30$~nb$^{-1}$/year for \PbXe\ interactions at \sqrtsNN = 72~GeV. A solid target is considered by ALICE, in conjunction with a beam splitting by a bent-crystal. In such a case, the beam halo is redirected by a crystal on a target inside the main ALICE magnet~\cite{Hadjidakis:2018JunePBC}; such a solution yields to $\int\calL \sim 5$~nb$^{-1}$/year for \PbA\ collisions at \sqrtsNN = 72~GeV. A comprehensive review of possible implementations can be found in~\cite{Hadjidakis:2018ifr}.

\section{The physics program}
The physics program of \AFTER\ covers three domains: heavy-ion collisions at the energy \sqrtsNN = 72 - 115~GeV~\cite{Trzeciak:2017csa},  studies of a structure of nuclear matter at the \textit{high-x} frontier, and investigations of the spin composition of a nucleon~\cite{Kikola:2017hnp}. While the \AFTER\ program includes many exciting opportunities, we briefly summarize here those relevant for quark-matter studies.
Experimental results on the elliptic flow and jet-quenching effects from the RHIC Beam Energy Scan (BES) program~\cite{Adamczyk:2017nof,Adamczyk:2013gw} suggest that such a partonic matter is created in the energy range of \AFTER. As such, one could study with \AFTER\ the QGP properties and the phase transition between the hadronic matter and the QGP. Two main assets facilitate such a research: the large heavy-quark yields expected in a single year of operation (both for open heavy flavor hadrons and heavy quarkonia) and a wide rapidity coverage.  Because of their large mass, the charm and bottom quarks are produced early in the nuclear collisions, during the initial interactions and with a large momentum transfer. Therefore, they carry information about all stages of the evolution of the created system. The total and differential production cross-section are well described by perturbative QCD calculations. Thus, they should constitute well-calibrated probes of both ``cold'' and ``hot'' nuclear matter. With \AFTER\, we expect precision studies of the charmonium, bottomonium and charmed meson production in \pp, \pA, and \AA\ interactions. The measurement of nuclear modification factors $R_{pA}$ for heavy flavor mesons and quarkonium states in \pA\ collisions is a powerful tool for constraining gluon nuclear parton distribution function (nPDF) in a nucleus~\cite{Kusina:2017gkz,Lansberg:2016deg} whose knowledge is essential to model in-medium interactions of heavy quarks. Figure~\ref{fig:gluon_RpA} shows the current uncertainty on the gluon nPDF in the Xe nucleus along with the expected one after the inclusion of \AFTER\ data. A drastic improvement is expected in the high-$x$ range. Such studies should however go along
with dedicated studies of the so-called intrinsic charm content~\cite{Brodsky:1980pb}, energy loss and/or nuclear absorption effects. Along the same lines it is possible to study the gluon and charm content of the deuteron~\cite{Brodsky:2018zdh}.

Regarding the \AA\ collisions, the feasibility studies done so far \cite{Trzeciak:2017csa,Massacrier:2015qba,Kikola:2015lka} show that nuclear modification factor $R_{AA}$ for \jpsi, $\psi(2S)$, $\Upsilon(nS)$ and D$^0$ will be measured with a percent-level precision. Such accurate data will allow for the effective use of quarkonium states as a QGP thermometer (with all the well known caveats of this approach). A scrupulous study of the charm-quark energy loss and collective behavior of charm quarks in the QGP will be possible by measurement of $R_{AA}$ and the elliptic flow $v_2$ as a function of transverse momentum and rapidity. In turn, the thermodynamic and transport properties of the QGP could be studied with an unprecedented precision. Besides, precise measurements of the charmonium and D-meson yields as a function of particle multiplicity in \pp\ and \pA\ collisions could shed new light on the collective behavior of partons in small systems.

\begin{figure}[hbt!]
\centering
\subfigure[~]{	\includegraphics[height=0.38\textwidth]{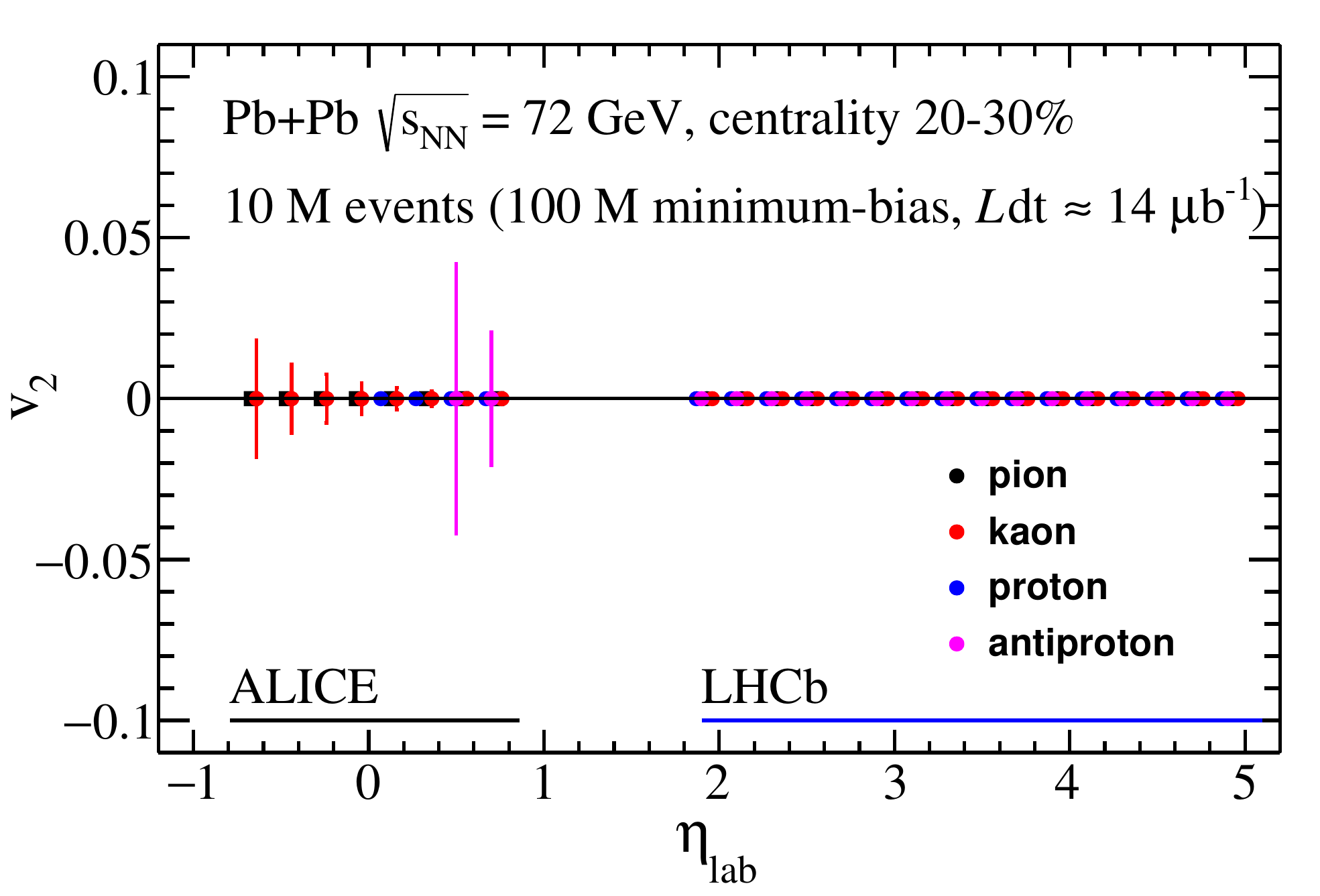}\label{fig:HI:flow}}
\subfigure[~]{	\includegraphics[height=0.37\textwidth]{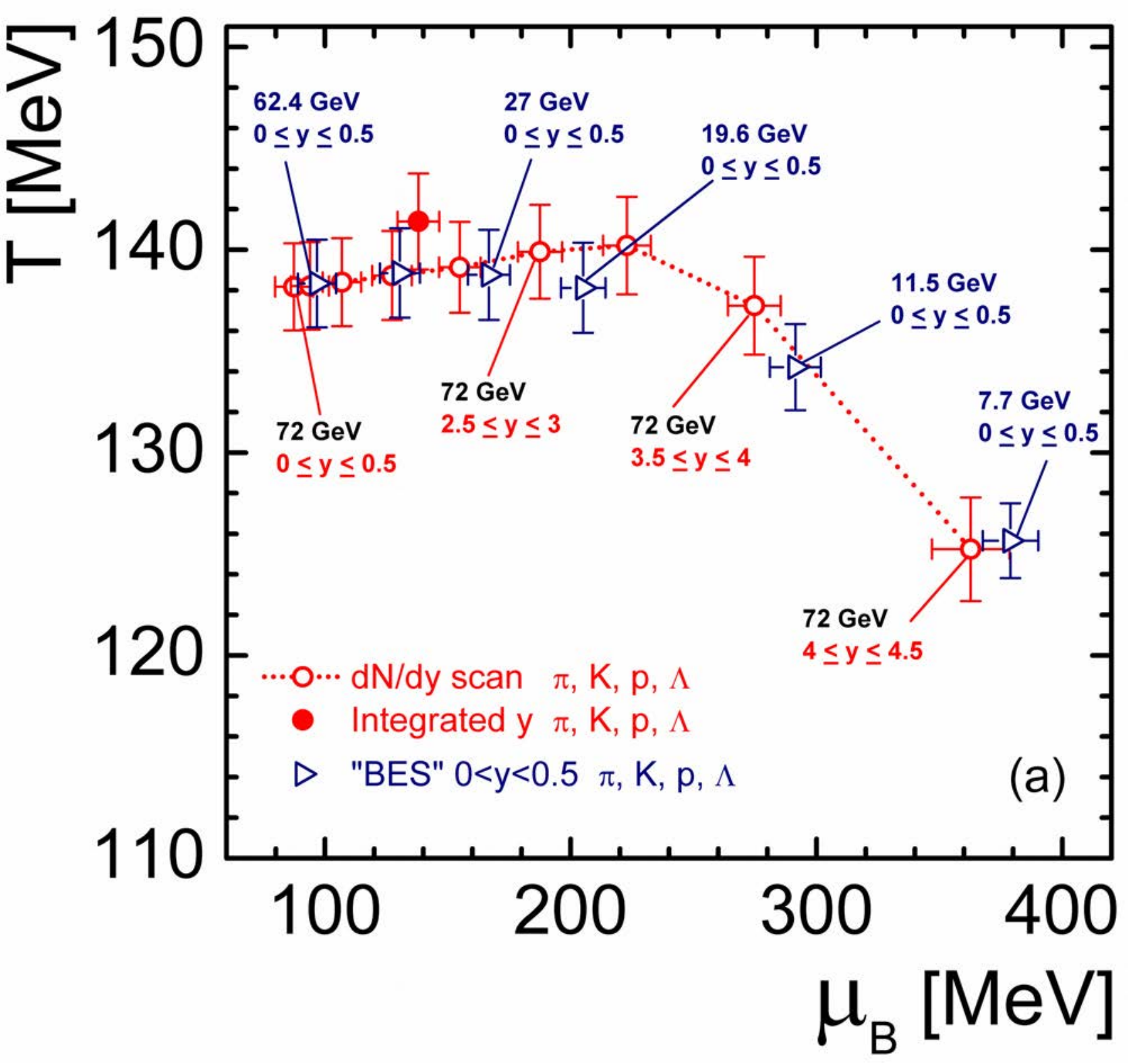}\label{fig:HI:muB}}
\caption{(a)  Statistical uncertainties on the measurement of the the $p_{T}$-integrated $v_2$ of identified hadrons as a function of the $\eta$ with the ALICE-like and LHCb-like detectors~\cite{Hadjidakis:2018ifr}. (b) HRG calculations for the range of temperature T and the chemical baryonic potential $\mu_B$ available in the top 10\% central \PbPb\ collisions at \sqrtsNN = 72 GeV~\cite{Begun:2018efg}. }
\label{fig:res-vs-rapidity}
\end{figure}

A large rapidity coverage is another useful instrument for the QGP study. The combined acceptance of the LHCb and ALICE detectors span over 7 units of pseudorapidity $\eta$. Measurements of the $v_2$ and yields over such a broad rapidity range will help to constrain the temperature dependence of the shear viscosity of the QGP. Figure \ref{fig:HI:flow} displays the expected statistical precision for identified hadron $v_2$ studies in \AFTER. Even with a small data sample (10$^6$ minimum-bias events) we expect highly accurate results, thus precise QGP property determinations. Varying the rapidity allows one to scan over the temperature T and the baryonic chemical potential $\mu_B$. Figure~\ref{fig:HI:muB} shows the $\mu_B$ vs. T dependence from the hadron resonance gas (HRG) model obtained by varying the rapidity interval at a fixed \sqrtsNN, or by performing a beam energy scan in a fixed rapidity bin. These results suggest that a rapidity scan done at \AFTER\ could give access to the similar $\mu_B$ range as in the RHIC BES program. Thus, the rapidity scan at \AFTER\ and the RHIC beam energy scan are complementary approaches to study the QCD phase diagram.

\section{Summary and current status}
In a recent review~\cite{Hadjidakis:2018ifr}, the \AFTER\ study group gathered compelling physics cases, implementation options and feasibility studies for a fixed-target program at the LHC. Such a program offers unique opportunities in the area of high-energy nuclear and particle physics and astrophysics. It is a possible extension of the LHC collider research program and it is one of the topics reviewed by the  Physics Beyond Collider working group (see \eg~\cite{Redaelli:2018IPAC}). Both the LHCb and the ALICE collaborations actively investigate the integration of fixed-target collisions within their experimental programs~\cite{Hadjidakis:2018JunePBC,Graziani:2018JunePBC,Maurice:2017iom}. We believe all these efforts will yield a wealth of remarkable physics results in the near future.


\end{document}